
\input phyzzx

%
%
\def\={\!=\!}
\def\+{\!+\!}
\def\-{\!-\!}
\def\g{\sqrt{G}}
\def\a{\alpha}
\def\b{\beta}
\def\ga{\gamma}
\def\d{\delta}
\def\e{\epsilon}
\def\half{{\textstyle{1\over2}}}
\def\quart{{\textstyle{1\over4}}}
\def\qh{{\textstyle{Q\over2}}}
\def\fourg{\sqrt{{}^{\scriptscriptstyle(4)}\!G}}
\def\fourR{{}^{\scriptscriptstyle(4)}\!R}
\def\ihi{\int_h^\infty}
\def\Vphi{{\textstyle{\partial V\over\partial\phi}}}
\def\Vpsi{{\textstyle{\partial V\over\partial\psi}}}
%
%
\Ref\related{ R. Marnelius \journal Nucl. Phys.&B211 (83) 14; \nextline
   C. Teitelboim \journal Phys. Lett.&B126 (83) 41; \nextline
   R. Jackiw, in \sl Quantum Theory of Gravity, \rm S. Christensen, ed.
   (Hilger, Bristol U.K. 1984); \nextline
   A.H. Chamseddine \journal Phys. Lett.&B256 (91) 379; \nextline
   T. Banks and M. O'Loughlin \journal Nucl. Phys.&B362 (91) 649; \nextline
   R.B. Mann, \sl ``Lower dimensional gravity", \rm Waterloo preprint
   WATPHYS-TH-91-07 (Oct. 1991), to appear in Proc. of 4th Canadian Conf. on
   General Relativity and Relativistic Astrophysics, Winnipeg, Canada,
   May 1991.}
\Ref\blackhole { E. Witten \journal Phys. Rev.&D44 (91) 314; \nextline
   S. Elitzur, A. Forge and E. Rabinovici
   \journal Nucl. Phys.&B359 (91) 581; \nextline
   G. Mandal, A. Sengupta and S. Wadia \journal Mod. Phys. Lett.&A6 (91) 1685.}
\Ref\nappi{ M.D. McGuigan, C.R. Nappi and S.A. Yost,
   \sl ``Charged Black Holes in Two-Dimensional String Theory,"
   \rm IAS preprint IASSNS-HEP-91/57 (Oct. 91) hep-th/9111038,
   to appear in \sl Nucl. Phys. \bf B\rm.}
\Ref\quantum{ J.G. Russo and A.A. Tseytlin, \sl ``Scalar-Tensor Quantum Gravity
   in Two Dimensions," \rm Stanford and Cambridge preprint SU-ITP-92-2
   DAMTP-1-1992 (Jan. 1992) hep-th/9201021; \nextline
   T.T. Burwick and A.H. Chamseddine, \sl ``Classical and Quantum
   Considerations of Two-dimensional Gravity," \rm Z\"urich preprint
   ZU-TH-4/92 (March 92) hep-th/9204002.}
\Ref\CFMP{ C.G. Callan, D. Friedan, E.J. Martinec and M.J. Perry
   \journal Nucl. Phys.&B262 (85) 593; \nextline
   E.S. Fradkin and A.A. Tseytlin \journal Nucl. Phys.&B261 (85) 1.}
\Ref\GHS{ D. Garfinkle, G.T. Horowitz and A. Strominger
   \journal Phys. Rev.&D43 (91) 3140.}
\Ref\horizons{ G.W. Gibbons and S.W. Hawking \journal Phys. Rev.&D15 (76)
2738.}
\Ref\CGHS{ C.G. Callan, S.B. Giddings, J.A. Harvey and A. Strominger
   \journal Phys. Rev.&D45 (92) 1005.}
\Ref\dealwis{ S.P. de Alwis, \sl ``Comments on No-Hair Theorems and Stability
   of Blackholes", \rm Boulder preprint COLO-HEP-267 (Jan. 1992)
   hep-th/9201053.}
\Ref\nohair{ J.E. Chase \journal Comm. Math. Phys.&19 (70) 276; \nextline
   J.D. Bekenstein \journal Phys. Rev.&D5 (72) 1239 \journal ibid.&D5 (72)
2403;
   \nextline C. Teitelboim \journal Phys. Rev.&D5 (72) 2941.}
\Ref\buchdahl{ H.A. Buchdahl \journal Phys. Rev.&115 (59) 1325; \nextline
   M. Cadoni \journal Phys. Rev.&D44 (91) 1115.}
%
%
\hfuzz=20pt
\nopubblock
\titlepage
\line{\hfil hep-th/9204026}
\line{\hfil IASSNS-HEP-92/22}
\line{\hfil March 1992}
{\bf\title{Dilaton Gravity and No-Hair Theorem in Two Dimensions\foot{ {\rm
Research supported in part by the Department of Energy,
contract DE-FG02-90ER40542, and by the Ambrose Monell Foundation}}}}
\author{Olaf Lechtenfeld\foot{
e-mail: \caps olaf@iassns.bitnet \rm or \caps
lechtenf@guinness.ias.edu}}
\andauthor{Chiara Nappi\foot{
e-mail: \caps nappi@iassns.bitnet}}
\address{School of Natural Sciences
\break Institute for Advanced Study
\break Princeton, NJ 08540}
\vfil
\abstract
We study a general class of
two-dimensional theories of the dilaton-gravity type inspired by string theory
and show that they admit charged multiple-horizon black holes.
These solutions are proved to satisfy scalar no-hair theorems.
\endpage
\pagenumber=1
\sequentialequations
Over the years many attempts have been made to formulate a non-trivial theory
of two-dimensional gravity (see, for example, ref.~[\related]).
More recently, attention has been
drawn to two-dimensional string-inspired models of gravity, which have been
shown to have black hole solutions~[\blackhole].
The non-triviality of these models arises from the non-minimal coupling of
a scalar field, the ``dilaton", to the scalar curvature.
In the presence of a
dilaton potential of the type produced by string loop corrections,
the black hole solutions may exhibit multiple horizons,
as discussed in~[\nappi],
where charged black holes in ``heterotic"
two-dimensional string theories were also studied.
In this letter we put together all these features and examine
a general class of two-dimensional string-inspired theories.
At the classical level,\foot{
For recent work on the quantum level, see~[\quantum].}
they all seem to provide
a consistent definition of two-dimensional gravity that has many analogies
with four-dimensional gravity.
We exploit these analogies to prove that generalized Reissner-Nordstr\o m
black-holes satisfy no-hair theorems.
This puts constraints on additional scalar fields
if they are to enter non-trivially in such solutions.

In the following we consider two-dimensional ``dilaton gravity"
conformally coupled to Maxwell and scalar fields,
$$
S_2\ =\ \int\! d^2x\;\g\,e^{-2\phi}\,\bigl[R + \ga(\nabla\phi)^2
- \quart e^{\e\phi}F^2 + \d(\nabla\psi)^2 + V(\phi,\psi)\bigr]
\eqn\Stwo $$
where $\g\equiv\sqrt{-\det g_{\mu\nu}}$, $\phi$ denotes the dilaton, and
$F_{\mu\nu}\=\partial_{[\mu}A_{\nu]}$ is the Maxwell field.
The scalar~$\psi$ is a spectator field, and the restriction to
a single scalar is merely for simplicity.
$S_2$ covers a large family of non-trivial actions for two-dimensional gravity.
For the special choice of
$\ga\=4$, $F\=\psi\=0$, and $V(\phi,0)\=-\Lambda$,
this is the usual bosonic string spacetime effective action~[\CFMP],
discussed in~[\blackhole,\nappi].
In addition, the Maxwell term (for $\e\=0$) appears for the heterotic string.
The cosmological constant~$\Lambda$ originates from the conformal anomaly.

In another specific case,
$\ga\=2$, $\e\=0$, and $V(\phi,\psi)\=2e^{2\phi}\+U(\psi)$,
our action is identical to the {\it four\/}-dimensional
$$
S_4\ =\ \int\!d^4x\;\fourg\,\bigl[
\fourR - \quart F^2 + \d(\nabla\psi)^2 + U(\psi)\bigr]
\eqn\Sfour $$
for spherically symmetric and purely electrically charged
field configurations,
as is obvious in the gauge
$$
{}^{\scriptscriptstyle(4)}\!ds^2\ =\
{}^{\scriptscriptstyle(2)}\!ds^2\ +\ e^{-2\phi(r)} d\Omega \quad,\quad
F\ =\ f(r)\,dr\wedge dt \quad.
\eqn\gaugefour $$
Here the ``dilaton" $\phi$ has become part of the four-dimensional metric.
Four-dimensional purely {\it magnetic\/} Maxwell fields,
$F=q\sin\theta d\theta\wedge d\phi$,
simply reduce to a modification $V\to V\-2q^2 e^{4\phi}$
of the potential $V(\phi,\psi)$ in two dimensions.
For example, all solutions of~[\GHS] can be obtained from our action~\Stwo\
with $\ga\=2,\,\d\=-2,\,F\=0$ and
$V(\phi,\psi)\=2e^{2\phi}\-2q^2 e^{4\phi-2a\psi}$.

Certain choices of the coefficients $\ga,\d,\e$ and the potential~$V$ in $S_2$
are related by Weyl rescaling.
Of course, conformal rescaling cannot remove the direct gravity-dilaton
coupling in~$S_2$, a unique feature of two-dimensional gravity.
However, $ds^2\to e^{\b\phi}ds^2$ does change the coefficients in~$S_2$,
$$
\ga\to\ga-\b^2\quad,\qquad\d\to\d\quad,\qquad\e\to\e-\b\quad,\qquad
V\to e^{\b\phi}V\quad,
\eqn\rescale $$
and can eliminate the dilaton kinetic term for~$\b\=\pm\sqrt{\ga}$.
In particular, for the traditional case of $\ga\=4$ and $V\=-\Lambda\+U$,
$\b\=2$ yields
$$
S_2^{\rm trad}\ \to\ \int\!d^2x\;\g\,\bigl[
e^{-2\phi}R -\Lambda - \quart e^{(\e-4)\phi}F^2
+ \d e^{-2\phi}(\nabla\psi)^2 + U(\psi)\bigr]\quad.
\eqn\Stwotrad $$
The classical spacetime generated by this action ($\ga\=0$) in the absence of
Maxwell and scalar fields is flat, but the dilaton solution still carries
information about the black hole singularity.
Had we chosen $V=-\Lambda e^{-2\phi}\+U(\psi)$ instead, another well-known
candidate action for two-dimensional gravity,
$\tilde S_2=\int\!d^2x\,\g e^{-2\phi}[R\-\Lambda\+\ldots]$,
would have emerged~[\related].
The transformation~\rescale\ implies that the distinction
between metric and dilaton is
largely an arbitrary one, since rescaling shifts structure from one to
the other.
Despite the simplicity of choosing $\ga\=0$, the analogy between
two-dimensional ``dilaton gravity" and four-dimensional pure gravity is best
pursued for nonzero~$\ga$.
The similarity between the two theories
will become even more evident in the gauge we will soon adopt.

The equations of motion for the general action $S_2$ are
$$
\eqalign{
0\ &=\ -2\nabla^2\phi+4(\nabla\phi)^2-V(\phi,\psi)-\quart e^{\e\phi}F^2 \crr
0\ &=\ R_{\mu\nu}+2\nabla_\mu\nabla_\nu\phi+
(\ga\-4)\bigl(\nabla_\mu\phi\nabla_\nu\phi
-g_{\mu\nu}(\nabla\phi)^2+\half g_{\mu\nu}\nabla^2\phi\bigr) \cr
&\qquad-\half e^{\e\phi}F_\mu^\rho F_{\nu\rho}
+{\textstyle{\e\over16}}g_{\mu\nu}e^{\e\phi}F^2
+\d\,\nabla_\mu\psi\nabla_\nu\psi -\quart g_{\mu\nu}\Vphi \crr
0\ &=\ \nabla^\mu\bigl(F_{\mu\nu}\,e^{(\e-2)\phi}\bigr) \crr
0\ &=\ -2\d\,\nabla^2\psi + 4\d\,\nabla\phi\cdot\nabla\psi + \Vpsi \cr}
\eqn\covmotion $$
where appropriate linear combinations have been formed.
Our interest focuses on static classical configurations
$\{F\=f(r)dr\wedge dt,\,\phi(r),\,\psi(r)\}$
with asymptotically flat metric,
$$
\eqalign{
ds^2\ &=\ -g(r)\, dt^2\ +\ g(r)^{-1} dr^2 \cr
&\to\ -dt^2+dr^2 \quad{\rm as}\quad r\to\infty}
\eqn\gauge $$
in the half-space ${\bf R}\!\times\!{\bf R}_+\ni\{t,r\}$.\foot{
Actually, this asymptotic behavior is too restrictive:
In two dimensions, a {\it linear\/} function~$g(r)$ can always be gauged
to~$g\=1$ by a suitable coordinate transfromation since $R\=-g''$.
Obviously, we should admit $g(r)\to A\+Br$ as $r\!\to\!\infty$.
In contrast,
a quadratic limit, $g\to A\+Br\-\Lambda r^2$, signifies a nonzero cosmological
constant (when $\ga\!\ne\!4$) and for $\Lambda\!>\!0$ leads to a cosmological
event horizon.
We do not consider such theories here~[\horizons].}
Horizons $r\=h_i$ are defined as zeros of~$g(r)$.
Note that our choice of gauge is not the conformal one but sets $\g\=1$
everywhere.\foot{
For a treatment of $S_2$ (without the Maxwell field) in conformal gauge
see~[\CGHS].}
The field variations~\covmotion\ in gauge~\gauge\ become
(primes mean~${d\over dr}$)
$$
\eqalign{
(g\phi')' - 2g\phi'^2 - \quart f^2 e^{\e\phi} + \half V(\phi,\psi)\ &=\ 0 \cr
2\phi'' + (\ga\-4)\phi'^2 + \d\psi'^2\ &=\ 0 \cr
(f\,e^{(\e-2)\phi})' \ &=\ 0 \cr
(g\psi')'-2g\phi'\psi'- {\textstyle{1\over2\d}}\Vpsi(\phi,\psi)\ &=\ 0\cr}
\eqn\motion $$
which already solves the Maxwell field, $f(r)=f_0\,e^{(2-\e)\phi(r)}$.

An analytic solution of~\motion\ is possible for a constant spectator field,
$\psi(r)\=\psi_0$, sitting at a potential minimum,
$\Vpsi(\phi,\psi_0)\=0$.
We abbreviate $V(\phi,\psi_0)=:V_0(\phi)$.
With this simplification we explicitely find
$$
\phi(r)\ =\ \cases{\phi_0-\qh r &for $\ga=4$,\cr
\phi_0+{2\over\ga-4}\ln(r\-r_0) &for $\ga\ne4$.\cr}
\eqn\phisol $$
$Q$, $r_0$ and $\phi_0$ are integration constants;
we shall set $r_0\=0$ without loss of generality.
The remaining equation results in
$$
g(r)\ =\ \cases{
e^{-Qr}\bigl[-2m+{1\over Q}\int^r \!dr'\, e^{Qr'}\,W(\phi(r'))\bigr]
&\qquad for $\ga=4$,\crr
r^{\a+1}\bigl[-2m-{1\over\a}\int^r \!dr'\, r'^{-\a}\,W(\phi(r'))\bigr]
&\qquad for $\ga\ne4$,\cr}
\eqn\gsol $$
with $\a\equiv{4\over\ga-4}$ and
$W(\phi)\equiv V_0(\phi)-\half f_0^2 e^{(4-2\e)\phi}$.
The integration constant~$m$ turns out to be the mass of the black
hole~[\blackhole,\nappi].
Demanding $g(r\!\to\!\infty)\to1$ restricts $\ga\le4$ ($\a\!\le\!0$),
$\e\le\ga/2$, and forces the dilaton potential to decay (in~$r$) as
$W\bigl(\phi(r\!\to\!\infty)\bigr)\to\a(\a\+1)e^{(4-\ga)(\phi-\phi_0)}$.
Obviously, a cosmological constant is allowed only for~$\ga\=4$ where
$W\bigl(\phi(r\!\to\!\infty)\bigr)\to Q^2\=-\Lambda$ is mandatory.

The type of dilaton potential to be expected from closed string loop
corrections is of the form
$$
V_{0,\rm string}(\phi)\ =\ \sum_{n\ge0} a_n\,e^{2n\phi}\quad.
\eqn\pot $$
Plugging this into eq.~\gsol, one gets
(setting $\e\=0$ for simplicity)
$$
g(r)\ =\ \cases{
{b_0\over Q^2}-2me^{-Qr}+{b_1\over Q}re^{-Qr}+
\sum_{n\ge2}{b_n\over(1-n)Q^2}\,e^{-nQr}
&for $\ga=4$,\crr
-2mr^{1+\a}-\sum_{n\ge0}{b_n\over n\a^2+\a(1-\a)}\,r^{2+n\a}
&for $\ga\ne4$,\cr}
\eqn\gsolstring $$
with $b_n=e^{2n\phi_0}(a_n\-\half\d_{n,2}f_0^2)$.
In case the above denominator vanishes, $n\a\-\a\+1\=0$, the contribution to
$g(r)$ is $-{b_n\over\a}r^{1+\a}\ln r$ instead of being singular.
Asymptotic flatness requires that
$\ga\=4\-2k$\quad($\a\=-{2\over k}$) with $k\=0,1,2\ldots$,
and $b_0,b_1,\ldots,b_{k-1}\=0$ with the possible exception of~$b_{k/2}$,
but $b_k$ must not vanish except for~$\ga\=0$.
Furthermore, we need $m\=0$ unless $\ga\=4,2,0$.
These metrics generalize the well-known Reissner-Nordstr\o m black hole,
as is most evident for $\ga\=2$ ($b_0\=0,\,b_1\=2$),
$$
g(r)\ =\ 1\ -\ 2m\,r^{-1}\ -\
{\textstyle{b_2\over2}r^{-2}\ -\ {b_3\over6}r^{-4}\ -\ \ldots}\quad.
\eqn\gsolRN $$
The emergence of multiple-horizon spacetimes only depends on the values
of the coefficients~$a_n$, which should be determined from string theory.
If we admit a non-zero cosmological constant, $\Lambda\=-a_0\=-b_0\!\ne\!0$,
the metric function $g(r)$ for $\ga\!\ne\!4$ will pick up
an additional term of~$-\Lambda r^2$
which leads to a cosmological event horizon for~$\Lambda\!>\!0$.
This is also true for the four-dimensional de~Sitter gravity~\Sfour\
which admits up to four horizons~[\horizons]
even though 4-d general covariance restricts
$V_0\=-\Lambda\+2e^{2\phi}$ in the equivalent
two-dimensional action.
For the two-dimensional case, examples
of interesting spacetime geometries associated with multiple horizons
have been discussed in~[\nappi].
Here we have generalized such solutions to the whole class~\Stwo\ of
dilaton-gravity-Maxwell theories.

The second issue addressed in this note is the question of
{\it no-hair theorems\/}.
In our point of view, the dilaton is integral part of two-dimensional gravity
since it is necessary for the existence of black holes in the first place.
Therefore such theorems should not apply to the dilaton~[\dealwis].
The question is rather whether our solutions to the
dilaton-gravity-Maxwell system can be
perturbed by a non-trivial static spectator field, $\psi'\!\ne0$.
We know that the {\it four\/}-dimensional Einstein-Maxwell theory obeys
no-hair theorems.
Given the analogy between 2-d dilaton gravity and 4-d pure gravity, it is
natural to ask whether our solutions~\gsolstring\ are also stabilized
by such theorems.
If the direct spectator-dilaton coupling in~\Stwo\
was absent this question would obviously be answered in the affirmative.
Due to the exponential coupling, however, the matter is less than trivial.
The main scope of this letter is to establish
conditions for the uniqueness of the above
solutions to two-dimensional ``dilaton gravity" in the presence of Maxwell
and scalar fields.

The  key observation is that the equation of motion for $\psi$
is the same as in four dimensions, since $\psi$
couples in~$S_2$ in exactly the same way as in~$S_4$ for a
spherically symmetric setting.
This will allow us to take over Bekenstein's proof of
the scalar no-hair theorem~[\nohair]:
The equation of motion for~$\psi$, last of eqs.~\motion, can be rewritten as
$$
\bigl(g\,e^{-2\phi}\,\psi'\bigr)'\ =\ {\textstyle{1\over2\d}}\,
e^{-2\phi}\,\Vpsi(\phi,\psi) \quad.
\eqn\psieq $$
We multiply \psieq\ by $(\psi\-\psi_0)$ and integrate
between the last horizon~$h$,
where $g(h)\=0$, and spatial infinity, where $g\!\to\!1$.
Integration by parts then leads to
$$
\ihi \bigl[(\psi\-\psi_0)\,g\,e^{-2\phi}\,\psi'\bigr]'\ =\
\ihi g\,e^{-2\phi}\,\psi'^2\ +\
{\textstyle{1\over2\d}}\ihi(\psi\-\psi_0)\,
e^{-2\phi}\,\Vpsi(\phi,\psi) \quad.
\eqn\bek $$
If the boundary term vanishes and both integrands on the right hand side are
non-negative, we may conclude that $\psi'\equiv0$ and thus $\psi\equiv\psi_0$.
For this to be true, however, four conditions have to be met:\nextline
a)~$\d>0$,\foot{
The solutions of~[\GHS] violate this condition.} \nextline
b)~regularity of $\psi$ and~$\phi$ at $r\=h$,\foot{
For a violation of this condition see [\buchdahl].} \nextline
c)~${\partial^2 V\over\partial\psi^2}\ge0$
in $[h,\infty]$ for any~$\phi$, \nextline
d)~$\psi\!\to\!\psi_0$ ``fast enough" as~$r\!\to\!\infty$.

The conditions above are the same as in four-dimensional gravity.
``Morally" speaking, the charged and multiple-horizon solutions to the
general class of theories in~\Stwo\ satisfy the classical no-hair theorem.
In practice, however,
the condition~d) above is more or less restrictive depending on the
asymptotic behaviour of the dilaton, {\it i.e.\/} on
the value of the coefficient~$\ga$.
It ensures the vanishing
of the boundary term and means that $e^{-2\phi}\psi'^2$ must decay
faster than~$1/r$.
Apparently, the fall-off assumption~d) is very restrictive for the traditional
dilaton case ($\ga\=4$).
To get an idea, one might try to neglect the back reaction of $\psi$
on our above solutions for~$\phi$ and estimate that $\psi$ should approach
$\psi_0$ faster than $e^{-{1\over2}Qr}/\sqrt{r}$ or $1/r^{1-\a/2}$ for
$\ga\=4$ or $\ga\!\ne\!4$, respectively.

Fortunately, we can employ a different strategy to substantially strengthen
the no-hair theorem for the {\it traditional\/} dilaton-gravity-Maxwell system.
When $\ga\=4$ and $\e\=0$ the gravitational and dilaton equations of motion
imply\foot{
The second of these equations is a linear combination of eqs.~\motion\ and
their derivatives.}
$$
\eqalign{
(g\phi')' - 2g\phi'^2\ &=\ \quart f^2 - \half V(\phi,\psi) \cr
g'' - 2\phi'g'\ &=\ f^2 - \half \Vphi(\phi,\psi) \cr
\phi''\ &=\ -\half\d\,\psi'^2 \ \le0 \cr}
\eqn\gravdileq $$
while the Maxwell solution is unchanged, $f=f_0\,e^{2\phi}$.
Since the dilaton exponential in~\bek\ is responsible for the weakness of the
standard argument, we trivially rewrite eq.~\bek\ as
$$
\ihi \bigl[(\psi\-\psi_0)\,g\,\psi'\bigr]'\ =\
\ihi g\,\psi'^2\ +\ {\textstyle{1\over2\d}}
\ihi(\psi\-\psi_0)\,\Vpsi(\phi,\psi)\ +\
2\ihi(\psi\-\psi_0)\,g\,\phi'\,\psi' \quad.
\eqn\bekone $$
The no-hair theorem may now be valid under the same conditions as above,
except for a much weaker fall-off constraint,
namely d')~$\psi\to\psi_0$ in an arbitrary fashion at large~$r$.
However, success hinges on the positivity of the last term in~\bekone.
A second partial integration produces
$$
2\ihi(\psi\-\psi_0)\,g\,\phi'\,\psi'\ =\
\ihi \bigl[(\psi\-\psi_0)^2\,g\,\phi'\bigr]'\ -\ \ihi(\psi\-\psi_0)^2(g\phi')'
\eqn\bektwo $$
which invites the application of the other classical equations,~\gravdileq,
$$
-(g\,\phi')'\ =\
\half\d\,g\psi'^2 - g'\phi'\ =\
\half\d\,g\psi'^2 - \half g'' + \half f_0^2e^{4\phi} - \quart \Vphi \quad.
\eqn\gphi $$

To proceed, obviously some information about sign and magnitude of $g'$
and~$\phi'$ in $[h,\infty]$ is needed.
To this end we restrict ourselves again to
$V(\phi,\psi)=Q^2\+U(\psi)$, \ie\ string tree level dilaton potential.
The first two of eqs.~\gravdileq\ can be turned into integral representations
for $\phi'$ and~$g'$,
$$
\eqalign{
\phi'(r)\ &=\ {e^{2\phi(r)}\over2g(r)}\,\int_h^r\!dr'\,\Bigl[
\half f_0^2 e^{2\phi(r')}
-e^{-2\phi(r')}\bigl(U(\psi(r'))+Q^2\bigr)\Bigr] \crr
g'(r)\ &=\ e^{2\phi(r)} \Bigl[c-f_0^2 \int_r^\infty\!dr'\,e^{2\phi(r')}
\Bigr] \quad, }
\eqn\intrep $$
with $c=2mQ>0$ from the large-$r$ limit.
The electrical charge~$q$ of the black hole hides in $\half f_0^2=q^2Q^2$.
Our asymptotic condition on~$g$ then translates to
$$
1\ =\ \ihi\!dr\,g'(r)\ =\ 2mQ\ihi\!dr\,e^{2\phi(r)}
-2q^2Q^2 \ihi\!dr\,e^{2\phi(r)}\int_r^\infty\!dr'\,e^{2\phi(r')}
\eqn\asy $$
which is a quadratic equation for $\int_h^\infty e^{2\phi}$ and solved by
$$
\ihi\!dr\,e^{2\phi(r)}\ =\
{1\over q^2Q}\bigl[{\textstyle m-\sqrt{m^2-q^2}}\bigr] \quad.
\eqn\intdil $$
Plugging this result back into eq.~\intrep\ we can bound $g'$ by
$$
g'(r)\ \ge\ e^{2\phi(r)}\;2Q\,{\textstyle\sqrt{m^2-q^2}}\ \ge0
\eqn\bound $$
outside the horizon!
We see that $g'$ may just vanish at the horizon in the extremal case, $|q|\=m$.
The sign of~$\phi'$, on the other hand, need not always be negative.
{}From its large-$r$ behavior and the fact that $\phi''\le0$ we conclude that
$\phi'$ is maximal for minimal~$r$, which on $[h,\infty]$ means~$r\=h$.
Applying l'H\^opital's rule to the first of eqs.~\intrep\ we arrive at
$$
\phi'(h)\ =\
{1\over2g'(h)}\Bigl[-U\bigl(\psi(h)\bigr)-Q^2+Q^2q^2 e^{4\phi(h)}\Bigr]
\eqn\phih $$
which is negative as long as
$$
U(\psi)\ \ge\ Q^2\bigl(q^2\,e^{4\phi}-1\bigr)
\eqn\Ucond $$
at the horizon.
For $U\ge0$ we may read this as a condition on the electric charge,
$$
|q|\ \le\ e^{-2\phi(h)} \quad.
\eqn\qcond $$
Looking back at eq.~\gphi\ we are able to complete the proof of the
no-hair theorem in the traditional $\ga\=4$, $\e\=0$, $\Vphi\=0$~case
for arbitrarily small decay of~$\psi\!\to\!\psi_0$,
provided the conditions a)--c) are
supplemented by e)~an upper bound on the electric charge of the black hole.

In summary, we have written down all static classical solutions to a general
class of two-dimensional dilaton-gravity-Maxwell theories and found them to
generically exhibit multiple-horizon spacetimes.
These solutions obey the classical no-hair theorem for scalar spectator fields.
For the standard string effective spacetime action
an improved argument does not rely on fall-off conditions for the spectator
field but requires a bound on the electric charge of the black hole.

\ack
We acknowledge discussions with J.~Schiff and E.~Weinberg.
\refout
\end